\documentclass[sigconf]{acmart}
\usepackage[T1]{fontenc}
\usepackage{graphicx}
\usepackage{tikz}
\usetikzlibrary{positioning, arrows.meta, shapes.geometric, calc, fit, backgrounds, shapes.multipart, decorations.pathreplacing}
\usepackage{bbding}
\usepackage[htt]{hyphenat}
\usepackage{subcaption}
\usepackage{xspace}
\usepackage{balance}
\usepackage{xcolor}    %
\usepackage{url}
\usepackage{hyperref}
\usepackage{cleveref}

\AtBeginDocument{%
  \providecommand\BibTeX{{%
    \normalfont B\kern-0.5em{\scshape i\kern-0.25em b}\kern-0.8em\TeX}}}

\copyrightyear{2026}
\acmYear{2026}
\setcopyright{cc}
\setcctype{by}
\acmConference[SCORED '26]{Conference on Software Supply Chain Offensive Research and Ecosystem Defenses}{October 06, 2026}{Prague, Czech Republic}
\acmBooktitle{Conference on Software Supply Chain Offensive Research and Ecosystem Defenses (SCORED '26), October 06, 2026, Prague, Czech Republic}
\acmDOI{10.1145/3848003.3848012}
\acmISBN{979-8-4007-3009-2/2026/10}

\author{Julien Malka}
\email{julien.malka@telecom-paris.fr}
\orcid{0009-0008-9845-6300}
\affiliation{\institution{LTCI, Télécom Paris, \\Institut Polytechnique de Paris}
 \city{Palaiseau}
 \country{France}
}
\author{Aman Sharma}
\email{amansha@kth.se}
\orcid{0000-0003-2263-7902}
\affiliation{\institution{KTH Royal Institute of Technology}
 \city{Stockholm}
 \country{Sweden}
}
\author{Martin Monperrus}
\email{monperrus@kth.se}
\orcid{0000-0003-3505-3383}
\affiliation{\institution{KTH Royal Institute of Technology}
 \city{Stockholm}
 \country{Sweden}
}
\author{Stefano Zacchiroli}
\email{stefano.zacchiroli@telecom-paris.fr}
\orcid{0000-0002-4576-136X}
\affiliation{\institution{LTCI, Télécom Paris, \\Institut Polytechnique de Paris}
 \city{Palaiseau}
 \country{France}
}
\author{Théo Zimmermann}
\email{theo.zimmermann@telecom-paris.fr}
\orcid{0000-0002-3580-8806}
\affiliation{\institution{LTCI, Télécom Paris, \\Institut Polytechnique de Paris}
 \city{Palaiseau}
 \country{France}
}

\newcommand{\stepnum}[1]{\raisebox{0.15ex}{\textcircled{\scriptsize #1}}}

\tikzset{
  txreg/.style={draw,align=center,font=\scriptsize,minimum height=4.3mm,text width=20mm,inner sep=1.5pt},
  txhdr/.style={txreg,fill=gray!12},
  txrx/.style={txreg,fill=orange!14},
  txrw/.style={txreg,fill=green!14},
  txmeta/.style={txreg,fill=gray!6},
  txadd/.style={txreg,fill=red!12,draw=red!65!black,thick},
  txaddt/.style={txreg,fill=blue!12,draw=blue!65!black,thick},
  txarr/.style={->,thick,>=Latex},
}

\newcommand{\chg}[1]{{\tiny\itshape\textcolor{red!70!black}{#1}}}
\tikzset{
  exreg/.style={draw,align=center,font=\scriptsize,minimum height=6.5mm,text width=33mm,inner sep=2pt},
  exhdr/.style={exreg,fill=gray!14},
  exseg/.style={exreg,fill=orange!16},
  exrw/.style={exreg,fill=green!16},
  exmeta/.style={exreg,fill=gray!7},
  exnote/.style={exreg,fill=blue!8},
  exadd/.style={exreg,fill=red!14,draw=red!65!black,thick},
  exaddt/.style={exreg,fill=blue!14,draw=blue!65!black,thick},
  exoff/.style={font=\tiny\ttfamily,anchor=east,text=black!55},
}

\tikzset{
  frow/.style={draw,align=center,font=\scriptsize,minimum height=6mm,text width=37mm,inner sep=1.5pt},
  fhdr/.style={frow,fill=gray!14},
  fph/.style={frow,fill=violet!10},
  fseg/.style={frow,fill=orange!16},
  fdat/.style={frow,fill=green!16},
  fmeta/.style={frow,fill=gray!8},
  fadd/.style={frow,fill=red!14,draw=red!65!black,thick},
  faddt/.style={frow,fill=blue!14,draw=blue!65!black,thick},
  imem/.style={draw,align=center,font=\scriptsize,minimum height=9mm,text width=27mm,inner sep=2pt},
  imemrx/.style={imem,fill=orange!16},
  imemrw/.style={imem,fill=green!16},
  imempay/.style={imem,fill=red!14,draw=red!65!black,thick},
}

\begin{document}

\title[]{Trusting-Trust Attack against an Entire Linux Distribution through Binary Manipulation}

\begin{abstract}
Ken Thompson's trusting-trust attack, in which a compromised compiler backdoors the programs it builds and reproduces the backdoor in subsequent rebuilds of itself, is often regarded as a threat specific to compilers.
We show that it is not.
We construct a complete trusting-trust attack around GNU \texttt{strip}, an ordinary build utility that neither inspects nor generates source code, using only manipulations of finished ELF files.
In one bootstrap execution of the NixOS Linux distribution, a single tampered \texttt{strip} in the binary seed implants its payload into each later \texttt{strip} rebuilt from unmodified source.
The infected \texttt{strip} in the final standard environment can then implant the payload into binaries of downstream packages, even though that environment has no runtime reference to the bootstrap seed.
On a real nixpkgs revision, the attack builds a complete graphical installer without failures and backdoors almost every one of its binaries, enabling arbitrary malicious behavior of the subverted packages.
\end{abstract}



\maketitle

\section{Introduction}

``Given enough eyeballs, all bugs are shallow.'' Eric S. Raymond~\cite{raymond2001cathedral} popularizes this statement as Linus's law in his essay on open source software development.
It means that making source code publicly available for review, including by security experts, makes free and open source software less likely to contain bugs, security vulnerabilities, or backdoors.
Yet software must first be compiled before it can run, which limits the trust that source availability alone provides. As Ken Thompson showed in his Turing Award lecture~\cite{thompson1984trusting}, some compromises remain invisible to source code review. In Thompson's construction, a malicious compiler recognizes both a security-sensitive program and the source of the compiler itself.  It inserts a backdoor into the former and reproduces this logic in the latter. The malicious source can then be removed: recompiling unmodified compiler source with the compromised binary propagates the compromise.

Thompson's attack, dubbed ``trusting trust'', was more than a thought experiment because he implemented a prototype.
However, few documented or studied implementations exist, and the practical persistence of such a backdoor remains unclear.
One difficulty is that the malicious compiler must recognize its own source code to produce an altered binary.
Significant changes to that source could therefore stop the backdoor from propagating without exposing it.

Although the trusting-trust attack is often regarded as a threat specific to compilers, we demonstrate how its underlying mechanism extends to other build utilities.
We instantiate the attack in GNU \texttt{strip}, a post-build utility from GNU binutils that removes information from object files and executables, in the context of the NixOS Linux distribution.
Because \texttt{strip} processes a large fraction of the executables produced by a distribution, compromising it provides an opportunity both to propagate the compromise and to affect downstream software.

We demonstrate an end-to-end attack in which a malicious \texttt{strip} binary placed in the initial NixOS bootstrap seed causes the compromise to reproduce in later \texttt{strip} binaries rebuilt from unmodified source, and subsequently modifies other executables processed during their build. Unlike Thompson's original construction, our attack operates entirely through ELF-level transformations and neither inspects nor modifies source code. It is therefore insensitive to source-level changes that would prevent a malicious compiler from recognizing its own source. To our knowledge, this is the first documented end-to-end implementation of a self-propagating trusting-trust attack carried by a non-compiling post-build utility through the bootstrap of a production Linux distribution.

This paper makes the following contributions:
\begin{itemize}
\item We show that the trusting-trust attack extends beyond compilers by designing a self-propagating compromise of GNU \texttt{strip} based entirely on ELF-level transformations that neither inspect nor modify source code (\Cref{sec:design}).
\item We demonstrate that the attack propagates end to end through one NixOS bootstrap execution and persists in the final standard environment even though its output has no runtime reference to the malicious seed (\Cref{sec:threat,sec:design}).
\item We evaluate the attack by rebuilding a large and representative package set from nixpkgs, measuring its effect on buildability, its reach across downstream executables, and its impact on their runtime behavior (\Cref{sec:evaluation}).
\end{itemize}

We release the complete implementation and experimental artifacts in our replication package~\cite{this-replication-package}.

\section{Background}
\label{sec:background}

\subsection{Bootstrapping Linux Distributions}
\label{sec:distributions}

A \emph{Linux distribution} is a curated collection of software (the kernel, a
compiler toolchain and C library, system services, and end-user
applications) assembled and shipped together as a complete operating system,
with a package manager that installs and updates the individual \emph{packages}.
Debian, Fedora, Arch Linux, and the Nix-based NixOS are such examples.
A distribution contains \emph{recipes} for building packages, each specifying the upstream source and the exact steps needed to turn it into an installable package.

A Linux distribution needs to be bootstrapped:  
producing packages requires a compiler, a linker, and a C library.
Construction of the entire distribution must
therefore begin from a small set of pre-existing binaries called the \emph{bootstrap seed}.  \emph{Bootstrapping} is the
process that starts from this seed and builds a build toolchain.
A distribution is self-hosted when it depends only on binary artifacts it
has built itself. 

\subsection{The Nixpkgs Bootstrap}
\label{sec:nix-bootstrap}

Nix is a functional package manager, a model in which packages are built from recipes (also called \emph{derivations}) written in a functional programming language. Derivations contain a precise definition of the build environment needed to build the packages and undeclared dependencies are excluded from builds by a sandboxing mechanism~\cite{dolstra2004nix}.

The package collection \emph{nixpkgs} gathers the package recipes from which the Linux distribution \emph{NixOS} is built.
Nixpkgs packages are built from a standard environment (or \texttt{stdenv}), containing a C toolchain and associated build helpers such as binutils, a shell, and basic Unix tools~\cite{nixpkgsManual}.
The standard environment also provides a \emph{generic builder} that runs each package through a fixed sequence of standard phases~\cite{nixpkgsManual}.
These phases unpack, configure, build, and install the sources.
A default post-install phase called \emph{fixup} then normalizes the installed outputs by invoking GNU \texttt{strip} on their binaries and adjusting their runtime paths.
Constructing \texttt{stdenv} poses the familiar chicken-and-egg bootstrap problem described in \Cref{sec:distributions}: those tools are needed to build themselves.
To solve it, Nixpkgs relies on a small set of pre-built binaries called the \emph{bootstrap seed} to build the standard environment from source.

\begin{figure}[t]
  \centering
  \begin{tikzpicture}[
      >=Latex,node distance=3.5mm,
      b/.style={draw,rounded corners,align=center,text width=56mm,
                minimum height=6.5mm,font=\small},
      fresh/.style={b,fill=blue!8,draw=blue!55!black,very thick},
      reuse/.style={b,fill=gray!8,draw=black!55,dashed}]
    \node[reuse] (s0) {\textbf{Bootstrap seed}\\Contains the prebuilt \texttt{strip}.};
    \node[fresh,below=of s0] (s1) {\textbf{Stage 1}\\Builds a new \texttt{strip}; the seed's \texttt{strip} processes it during fixup.};
    \node[reuse,below=of s1] (mid) {\textbf{Intermediate stages}\\Rebuild GCC and libc while retaining the stage-1 \texttt{strip}.};
    \node[fresh,below=of mid] (s4) {\textbf{Stage 4}\\Builds another \texttt{strip}; the retained \texttt{strip} processes it during fixup.};
    \node[reuse,below=of s4] (fin) {\textbf{Final standard environment}\\Contains the rebuilt \texttt{strip}, with no runtime reference to the seed.};
    \draw[->,thick] (s0) -- (s1);
    \draw[->,thick] (s1) -- (mid);
    \draw[->,thick] (mid) -- (s4);
    \draw[->,thick] (s4) -- (fin);
  \end{tikzpicture}
  \caption{Relevant portion of the evaluated nixpkgs bootstrap. Arrows show the order of the stages. Heavy-bordered boxes mark stages that build \texttt{strip} from source as part of binutils; dashed boxes contain or retain an existing copy. Each newly built \texttt{strip} is processed by a previous \texttt{strip} during fixup.}
  \label{fig:bootstrap}
\end{figure}
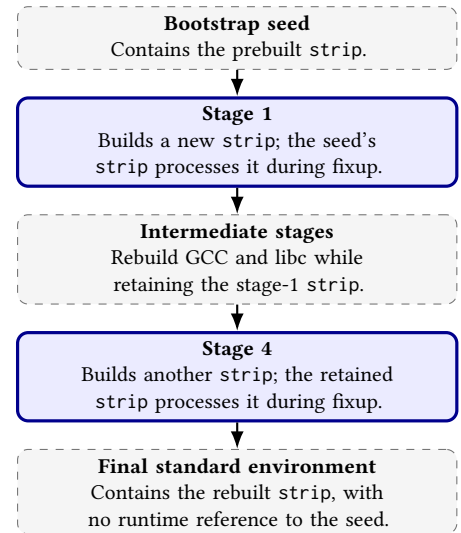

The seed bundles roughly twenty prebuilt programs, including a C compiler, binutils, the C library, and core utilities such as \texttt{bash}, \texttt{coreutils}, \texttt{tar}, \texttt{gzip}, and \texttt{patch}~\cite{trofiNixpkgsBootstrap}.
The bootstrap rebuilds the toolchain and basic utilities from source in several stages, each using tools produced by earlier stages.
The final \texttt{stdenv} has no runtime reference to the bootstrap seed, even though tools derived from the seed were used to build it.
Binutils, and with it \texttt{strip}, is built more than once to remove these runtime references.
The first source-built \texttt{strip} still uses the seed's compiler, while a later bootstrap stage builds \texttt{strip} with the rebuilt toolchain.
\Cref{fig:bootstrap} shows the relevant portion of this sequence.

\subsection{Trusting Trust Attack}

The original trusting-trust attack has been described by Thompson~\cite{thompson1984trusting}.  
In this attack, a compromised compiler binary builds a new compiler from unmodified source. That source contains no malicious logic. During compilation, however, the compromised binary injects malicious logic into the new binary. Because the injection happens during compilation rather than in the source, inspecting the source code cannot detect it.
The injected logic includes a copy of itself. When the newly built compiler later compiles another copy of the compiler, it performs the same injection. The attack propagates this way indefinitely, regardless of how many times the source is audited.

\subsection{Linux Internals}

\subsubsection{The \texttt{strip} utility}

GNU \texttt{strip}, part of binutils, removes symbols and other information
from object files.  \texttt{strip} is the tool the generic builder runs in the fixup phase~\cite{nixpkgsManual}.  Thus \texttt{strip} is routinely granted write access to executable package outputs.

\subsubsection{ELF Format}
\label{sec:elf-background}

The Executable and Linkable Format (ELF) is the standard binary format for Linux executables, shared libraries, and object files.
It is consumed both by the operating system to load and execute a program and by build tools to transform programs.

\begin{figure*}[t]
  \centering
  \begin{tikzpicture}[
      row/.style={draw,align=center,font=\scriptsize,minimum height=5mm,text width=48mm,inner sep=1.5pt},
      hdr/.style={row,fill=gray!14},
      phL/.style={row,fill=orange!16,minimum height=4.8mm},
      phLd/.style={row,fill=green!16,minimum height=4.8mm},
      phD/.style={row,fill=gray!8,minimum height=4.8mm,text=black!65},
      seg/.style={row,fill=orange!16},
      rw/.style={row,fill=green!16},
      note/.style={row,fill=blue!8},
      meta/.style={row,fill=gray!7},
      mem/.style={draw,align=center,font=\scriptsize,minimum height=9mm,text width=30mm,inner sep=2pt},
      memrx/.style={mem,fill=orange!16},
      memrw/.style={mem,fill=green!16},
      off/.style={font=\tiny\ttfamily,anchor=east,text=black!55},
      tick/.style={black!45,line width=0.3pt},
      maparr/.style={->,thick,>=Latex}]
    \node[hdr] (H) {ELF header\\{\tiny\ttfamily e\_entry=0x401050}\\{\tiny\ttfamily e\_phoff=0x0040 \textbullet\ e\_shoff=0x3400}};
    \node[phL,below=0pt of H]  (l1) {\texttt{PT\_LOAD} \ R\,E \hfill {\tiny file 0x1050 \textbullet\ vaddr 0x401050}};
    \node[phLd,below=0pt of l1] (l2) {\texttt{PT\_LOAD} \ R\,W \hfill {\tiny file 0x1200 \textbullet\ vaddr 0x402200}};
    \node[phD,below=0pt of l2] (pi) {\texttt{PT\_INTERP} \hfill {\tiny names ld-linux, not mapped}};
    \node[phD,below=0pt of pi] (pn) {\texttt{PT\_NOTE} \hfill {\tiny build-id note, not mapped}};
    \node[note,below=0pt of pn] (N) {\texttt{.note.gnu.build-id}};
    \node[seg,below=0pt of N] (T) {\texttt{.text}};
    \node[rw,below=0pt of T] (D) {\texttt{.data}};
    \node[meta,below=0pt of D] (S) {section headers};
    \node[draw=violet!65!black,very thick,rounded corners=2pt,fit=(l1)(l2)(pi)(pn),inner sep=1pt] (phbox) {};
    \node[font=\scriptsize\bfseries,text=violet!65!black,rotate=90,anchor=south]
      at ([xshift=-14mm]phbox.west) {program headers};
    \foreach \n/\v in {H/0x0000, l1/0x0040, N/0x02c4, T/0x1050, D/0x1200, S/0x3400}{
      \node[off] at ([xshift=-2.5mm]\n.north west) {\v};
      \draw[tick] ([xshift=-1.6mm]\n.north west) -- (\n.north west);
    }
    \node[font=\scriptsize\bfseries,align=center,above=1mm of H] {ELF file};
    \node[memrw,right=46mm of l1,yshift=-2mm] (MRW) {RW \texttt{PT\_LOAD}\\{\tiny vaddr 0x402200}\\{\tiny \texttt{.data} \textbullet\ R\,W}};
    \node[font=\tiny,below=6mm of MRW] (gap) {unmapped gap};
    \node[memrx,below=6mm of gap] (MRX) {RX \texttt{PT\_LOAD}\\{\tiny vaddr 0x401050}\\{\tiny \texttt{.text} \textbullet\ R\,E}};
    \node[font=\scriptsize\bfseries,above=1mm of MRW] {virtual memory};
    \draw[maparr,orange!75!black] (l1.east) to[out=0,in=180]
      node[pos=0.9,above,font=\tiny]{maps} (MRX.west);
    \draw[maparr,green!55!black] (l2.east) to[out=0,in=180]
      node[pos=0.85,above,font=\tiny]{maps} (MRW.west);
    \draw[->,thick,red!70!black] (H.east) to[out=0,in=150]
      node[pos=0.72,above,font=\tiny,fill=white,inner sep=0.8pt]{\texttt{e\_entry}} (MRX.north west);
  \end{tikzpicture}
  \caption{Simplified example of an ELF file and associated memory layout when running it. The program header table has four entries. The two \texttt{PT\_LOAD} entries create mapped regions.  Each \texttt{PT\_LOAD} identifies its bytes by a file offset and declares the virtual address to map them at. \texttt{PT\_INTERP} and \texttt{PT\_NOTE} are metadata and map no new memory. Execution starts at \texttt{e\_entry} inside the executable segment.}
  \label{fig:elf-memory}
\end{figure*}
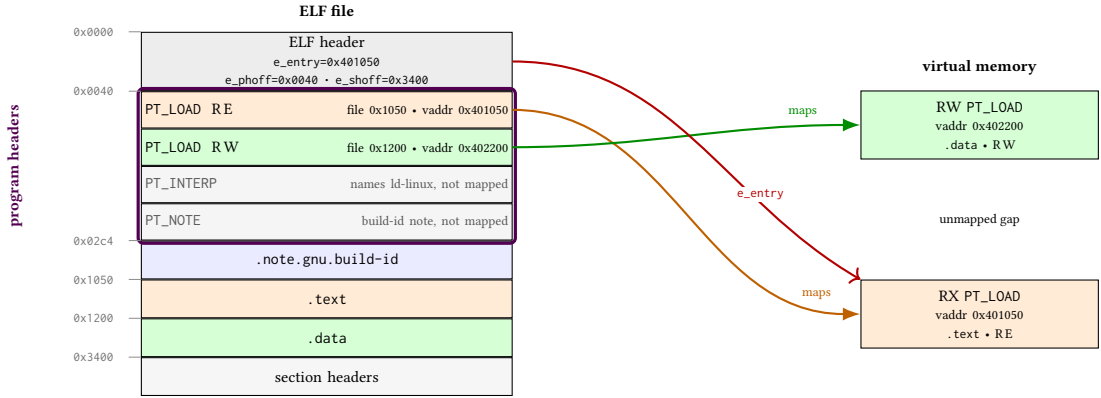

\Cref{fig:elf-memory} presents a running example of a simplified ELF file and how it is mapped in memory by the operating system loader. An ELF file begins with a header that identifies the target architecture and file type and locates two
different tables: the \emph{program header table} (located by the \texttt{e\_phoff} entry) and the \emph{section header table} (located by the \texttt{e\_shoff} entry).  The header also contains \texttt{e\_entry}, the virtual address at which execution begins~\cite{sysvElfAbi}.

The program header table tells the operating-system loader how to load the executable.
Each entry has a type that determines how the loader interprets it.
A \texttt{PT\_LOAD} entry describes a loadable \emph{segment}, specifying its file offset, virtual address, size, alignment, and permissions.
Other entry types provide auxiliary information.
In particular, \texttt{PT\_INTERP} specifies the dynamic loader, and \texttt{PT\_NOTE} identifies notes such as ABI or build metadata.
Once the loadable segments are mapped, the kernel transfers control to the \texttt{e\_entry} address.  It does not need the section table to start the program.

Section headers provide a complementary, tool-oriented view of many of the same bytes.  Named sections such as \texttt{.text}, \texttt{.data}, symbol tables, and note sections describe logical units used by linkers, debuggers, and binary utilities.  A segment may contain several sections, and not all sections need to be mapped at runtime. GNU \texttt{strip} reasons about this representation when deciding which information to retain or discard.

Program headers and section headers serve different consumers, but they must remain coherent with one another.

\subsubsection{ELF transformations}
\label{sec:elf-algebra}

The attack described in this paper does not act on the source code of the packages it targets.
It manipulates the binary files directly. We call these binary-level edits \emph{ELF transformations}~\cite{cesare1998unix}.

We use three transformations defined below that add regions and adjust metadata. They do not rewrite existing program code or data.

\begin{itemize}
\item \emph{Append} adds new bytes at the end of the file;
\item \emph{Repurpose} hijacks one spare metadata slot, a \texttt{PT\_NOTE} program header the loader does not need for execution, to use as an executable segment instead;
\item \emph{Redirect} changes a few ELF-header fields so that the loader and the binary tools notice what was appended.
\end{itemize}

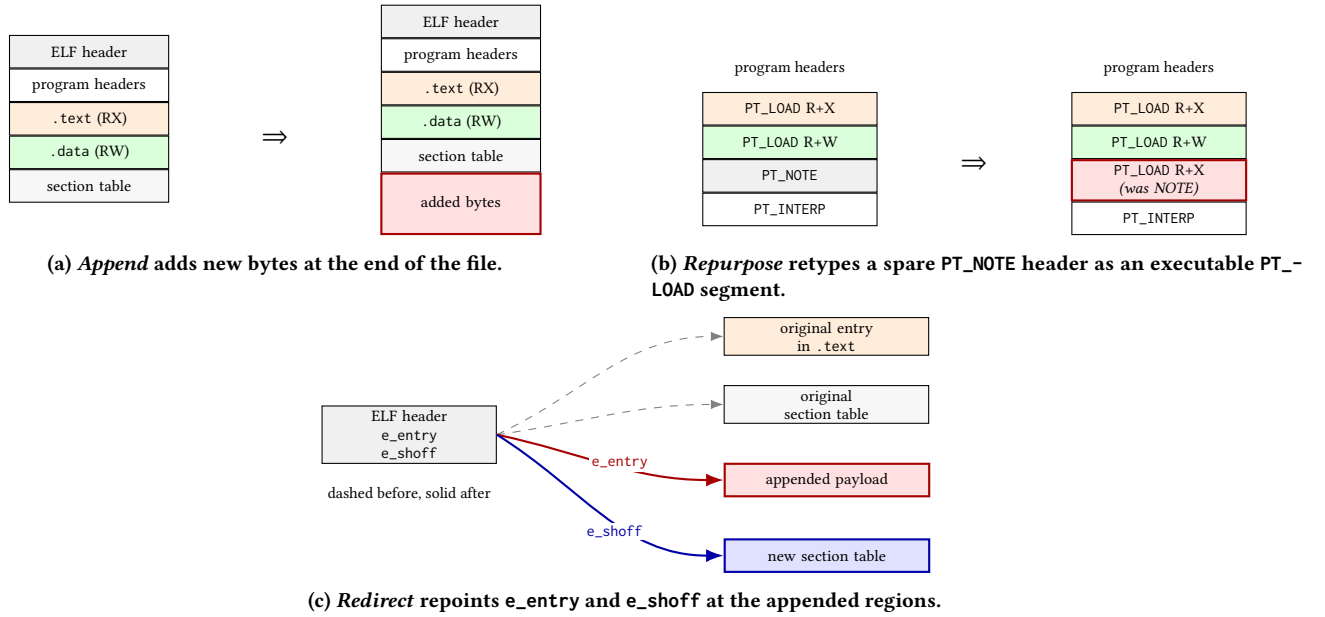
\begin{figure*}[t]
  \centering
  \begin{subfigure}[t]{0.48\textwidth}
    \centering
    \begin{tikzpicture}
      \node[txhdr,yshift=-4mm] (ch) {ELF header};
      \node[txreg,below=0pt of ch] (cp) {program headers};
      \node[txrx,below=0pt of cp] (ct) {\texttt{.text} (RX)};
      \node[txrw,below=0pt of ct] (cd) {\texttt{.data} (RW)};
      \node[txmeta,below=0pt of cd] (cs) {section table};
      \node[txhdr,right=28mm of ch,yshift=4mm] (ch2) {ELF header};
      \node[txreg,below=0pt of ch2] (cp2) {program headers};
      \node[txrx,below=0pt of cp2] (ct2) {\texttt{.text} (RX)};
      \node[txrw,below=0pt of ct2] (cd2) {\texttt{.data} (RW)};
      \node[txmeta,below=0pt of cd2] (cs2) {section table};
      \node[txadd,below=0pt of cs2,minimum height=8mm] (cx2) {added bytes};
      \node[font=\large] at ($(cd.east)!0.5!(cd2.west)+(0,0)$) {$\Rightarrow$};
    \end{tikzpicture}
    \caption{\emph{Append} adds new bytes at the end of the file.}
    \label{fig:tr-append}
  \end{subfigure}
  \hfill
  \begin{subfigure}[t]{0.48\textwidth}
    \centering
    \begin{tikzpicture}
      \node[font=\scriptsize] (t1) {program headers};
      \node[txrx,text width=22mm,below=1mm of t1] (e1) {\texttt{PT\_LOAD} R+X};
      \node[txrw,text width=22mm,below=0pt of e1] (e2) {\texttt{PT\_LOAD} R+W};
      \node[txreg,text width=22mm,below=0pt of e2,fill=gray!12] (e3) {\texttt{PT\_NOTE}};
      \node[txreg,text width=22mm,below=0pt of e3] (e4) {\texttt{PT\_INTERP}};
      \node[font=\scriptsize,right=32mm of t1] (t2) {program headers};
      \node[txrx,text width=22mm,below=1mm of t2] (f1) {\texttt{PT\_LOAD} R+X};
      \node[txrw,text width=22mm,below=0pt of f1] (f2) {\texttt{PT\_LOAD} R+W};
      \node[txadd,text width=22mm,below=0pt of f2] (f3) {\texttt{PT\_LOAD} R+X\\{\scriptsize\emph{(was NOTE)}}};
      \node[txreg,text width=22mm,below=0pt of f3] (f4) {\texttt{PT\_INTERP}};
      \node[font=\large] at ($(e3.east)!0.5!(f3.west)+(0,2mm)$) {$\Rightarrow$};
    \end{tikzpicture}
    \caption{\emph{Repurpose} retypes a spare \texttt{PT\_NOTE} header as an
      executable \texttt{PT\_LOAD} segment.}
    \label{fig:tr-repurpose}
  \end{subfigure}

  \medskip
  \begin{subfigure}[t]{0.62\textwidth}
    \centering
    \begin{tikzpicture}
      \node[txhdr,text width=22mm] (H) {ELF header\\\texttt{e\_entry}\\\texttt{e\_shoff}};
      \node[txrx,text width=26mm,right=30mm of H,yshift=13mm] (txt) {original entry\\in \texttt{.text}};
      \node[txmeta,text width=26mm,right=30mm of H,yshift=4mm] (os) {original\\section table};
      \node[txadd,text width=26mm,right=30mm of H,yshift=-6mm] (pl) {appended payload};
      \node[txaddt,text width=26mm,right=30mm of H,yshift=-16mm] (ns) {new section table};
      \draw[->,dashed,gray,>=Latex] (H.east) to[out=25,in=180] (txt.west);
      \draw[->,dashed,gray,>=Latex] (H.east) to[out=8,in=180] (os.west);
      \draw[txarr,red!65!black] (H.east) to[out=-12,in=180]
        node[above,font=\scriptsize,pos=0.55,yshift=-2pt,fill=white,inner sep=0.8pt]{\texttt{e\_entry}} (pl.west);
      \draw[txarr,blue!65!black] (H.east) to[out=-32,in=180]
        node[below,font=\scriptsize,pos=0.55,fill=white,inner sep=0.8pt]{\texttt{e\_shoff}} (ns.west);
      \node[font=\scriptsize,align=left,below=2mm of H]
        {dashed before, solid after};
    \end{tikzpicture}
    \caption{\emph{Redirect} repoints \texttt{e\_entry} and \texttt{e\_shoff} at
      the appended regions.}
    \label{fig:tr-redirect}
  \end{subfigure}
  \caption{The three ELF transformations we define, each shown as a before/after
  edit to the file in the style of Figure~\ref{fig:elf-memory}.}
  \label{fig:transformations}
\end{figure*}

These transformations preserve the existing program code and data, although repurpose and redirect modify ELF metadata.
Figure~\ref{fig:transformations} shows each transformation as a concrete before/after edit to the file.

\section{Threat Model}
\label{sec:threat}

We assume the attacker wants to attack a Linux distribution by targeting its binary bootstrap seed.
The attacker's capabilities are as follows.
The attacker can replace one executable in the bootstrap seed of the distribution.
The attacker cannot modify any distribution recipe or package sources.  

This capability models compromise before the seed is included in the distribution: for example,
by compromising the system that produces or publishes bootstrap archives, or by making an authorized update to the pinned seed, or with social engineering of the process by which maintainers decide and merge a new seed.

\section{The Bootstrapping Attack}
\label{sec:design}

\subsection{Attack Objective}
\label{sec:objective}

The objective of the attack is to compromise every downstream user of the distribution, and to keep doing so indefinitely.
We design and implement the attack through the \texttt{strip} binary, because it touches all binaries in a distribution.
Unlike a compiler such as \texttt{gcc}, which affects only the programs written in the languages it compiles, \texttt{strip} operates on finished ELF binaries directly and therefore reaches the outputs of every language and toolchain.

So, per the trusting trust attack model, a single tampered seed \texttt{strip} can both
(1) infect any application target and
(2) reproduce the compromise in the next built \texttt{strip}.
The attack is invisible to an auditor who reads the sources or the recipe.

\begin{figure*}[t]
  \centering
  \includegraphics[width=0.9\textwidth]{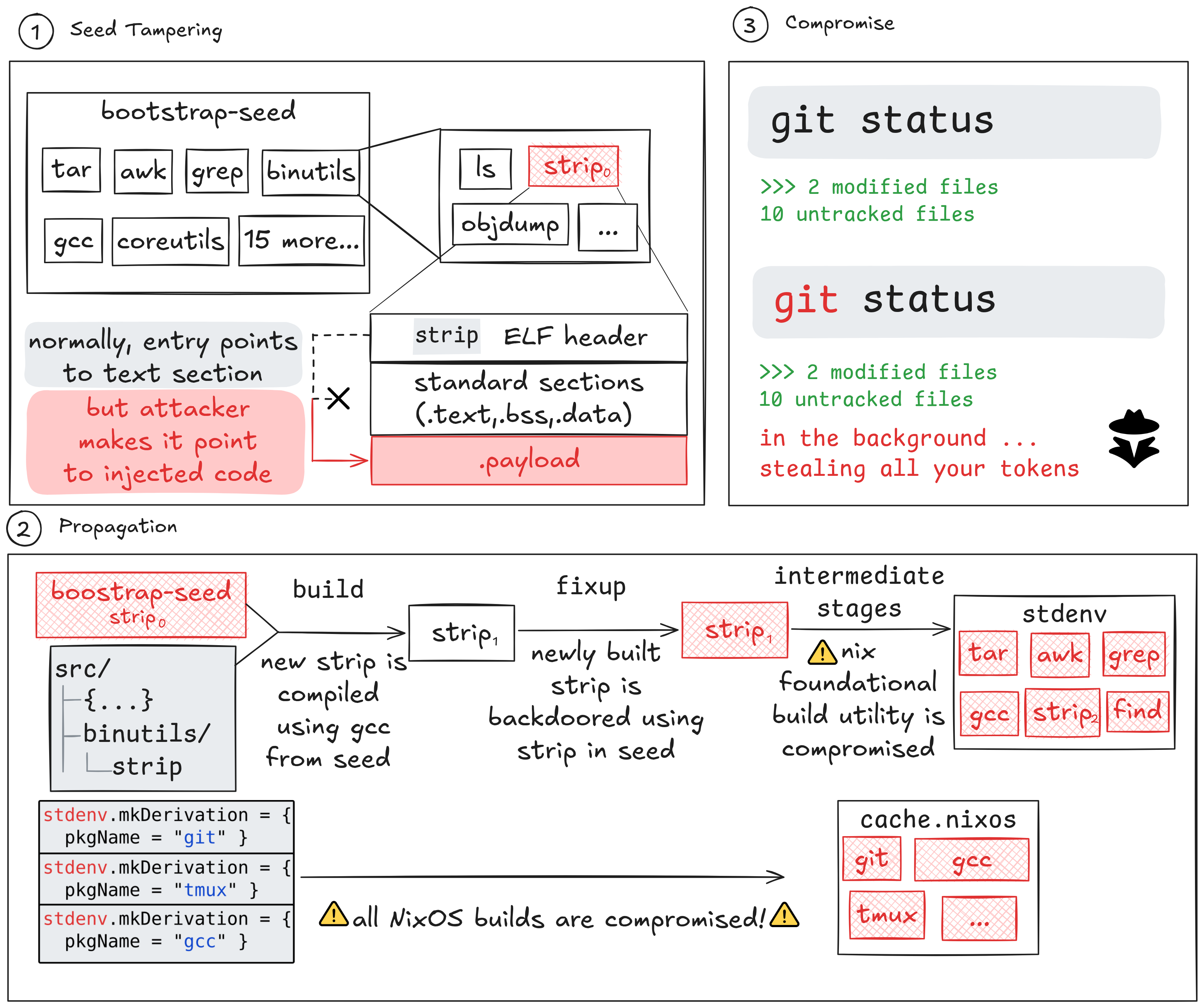}
  \caption{Overview of the attack design and implementation.}
  \label{fig:overview}
\end{figure*}

\subsection{Attack Overview}
\label{sec:overview}

Figure~\ref{fig:overview} summarizes the attack in three stages.
\emph{Seed tampering} builds a trojaned seed containing a compromised \texttt{strip}.
\emph{Propagation} carries that malicious behavior across the NixOS bootstrap to every later \texttt{strip}.
\emph{Compromise} occurs when a subverted application executable runs on the victim's machine.

Seed tampering (\Cref{sec:seed-tampering}) constructs the trojaned seed.
Propagation (\Cref{sec:propagation}) spreads it through the bootstrap.
Compromise (\Cref{sec:compromise}) describes the observable effect on the victim.

\subsection{Seed Tampering}
\label{sec:seed-tampering}
\label{sec:pattern}

Our attack instantiates \texttt{strip} as a \emph{self-propagating build utility}.
An earlier copy of such a utility processes a newly built copy, which later assumes the same role in the bootstrap.
Three conditions make this pattern possible, and \texttt{strip} satisfies all of them.
First (C1), the bootstrap must trust the utility before it builds the utility from source, and \texttt{strip} ships in the bootstrap seed.
Second (C2), an earlier copy must modify a newly built copy, and \texttt{strip} rewrites binaries during the fixup phase.
Third (C3), the modified copy must later perform the same operation, and every later stage runs \texttt{strip} again for fixup.
We call this propagation path from one copy to the next the \emph{successor edge}.

We build the trojaned \texttt{strip} by injecting the payload into the ELF file of the genuine \texttt{strip} with the ELF transformations of Section~\ref{sec:elf-algebra}.

The injector \emph{appends} the payload at the end of the ELF file.
This has no effect on execution at that point, because the region is not mapped in memory, as it is not referenced by the program headers.
The injector then \emph{repurposes} a non-functional \texttt{PT\_NOTE} program header and turns it into a \texttt{PT\_LOAD} entry that points at the newly added payload, so that the operating system loads it as an executable memory segment.
This step assumes the target carries at least one spare \texttt{PT\_NOTE} header, which holds for the toolchain binaries we target because the linker emits a build-id note (\texttt{.note.gnu.build-id}) by default.
Finally, the injector \emph{redirects} normal execution by modifying the \texttt{e\_entry} ELF header, so that the kernel jumps directly to the payload when it executes the program.
The first stage of Figure~\ref{fig:overview} shows this redirection, where the entry point that normally reaches the \texttt{.text} section is made to point at the injected code instead.
The payload also embeds the original \texttt{e\_entry} so that it can return to normal execution after performing its actions.

\begin{figure*}[t]
  \centering
  \begin{subfigure}[t]{0.29\textwidth}
    \centering
    \begin{tikzpicture}
      \node[fhdr] (H) {ELF header\\{\tiny\ttfamily e\_entry=0x401050 \textbullet\ e\_shoff=0x3400}};
      \node[fph,below=0pt of H] (P) {program headers\\{\tiny \texttt{PT\_LOAD}$\times$2, \texttt{PT\_INTERP}, \texttt{PT\_NOTE}}};
      \node[fseg,below=0pt of P] (T) {\texttt{.text} \hfill {\tiny R\,E}};
      \node[fdat,below=0pt of T] (D) {\texttt{.data} \hfill {\tiny R\,W}};
      \node[fmeta,below=0pt of D] (S) {section headers \hfill {\tiny off 0x3400}};
    \end{tikzpicture}
    \caption{The finished file, before injection.}
    \label{fig:elf-before}
  \end{subfigure}
  \hfill
  \begin{subfigure}[t]{0.37\textwidth}
    \centering
    \begin{tikzpicture}
      \node[fhdr] (H) {ELF header \ \chg{redirect}\\{\tiny\ttfamily e\_entry=\textcolor{red!70!black}{0x403000} \textbullet\ e\_shoff=\textcolor{red!70!black}{0x4160}}};
      \node[fph,below=0pt of H] (P) {program headers \ \chg{repurpose}\\{\tiny \texttt{PT\_NOTE}$\to$\texttt{PT\_LOAD} \textcolor{red!70!black}{0x403000} R\,E}};
      \node[fseg,below=0pt of P] (T) {\texttt{.text} \hfill {\tiny unchanged}};
      \node[fdat,below=0pt of T] (D) {\texttt{.data} \hfill {\tiny unchanged}};
      \node[fmeta,below=0pt of D] (S) {old section headers \hfill {\tiny bypassed}};
      \node[fadd,below=0pt of S] (X) {payload \textbullet\ \texttt{.payload} \ \chg{append}\\{\tiny off 0x4000 \textbullet\ vaddr 0x403000}};
      \node[faddt,below=0pt of X] (NT) {new section headers \ \chg{append}\\{\tiny off 0x4160}};
    \end{tikzpicture}
    \caption{After injection, two ELF header fields change, one program header is repurposed, and two regions are appended. The original program code and data are preserved.}
    \label{fig:elf-after}
  \end{subfigure}
  \hfill
  \begin{subfigure}[t]{0.32\textwidth}
    \centering
    \begin{tikzpicture}
      \node[imempay] (mpay) {RX \texttt{PT\_LOAD}\\{\tiny 0x403000 \textbullet\ payload}};
      \node[font=\tiny,below=5mm of mpay] (g1) {unmapped gap};
      \node[imemrw,below=5mm of g1] (mrw) {RW \texttt{PT\_LOAD}\\{\tiny 0x402200 \textbullet\ \texttt{.data}}};
      \node[imemrx,below=1.5mm of mrw] (mrx) {RX \texttt{PT\_LOAD}\\{\tiny 0x401050 \texttt{.text}}\\{\tiny orig entry 0x401050}};
      \node[font=\scriptsize\bfseries,above=6mm of mpay] {higher addresses $\uparrow$};
      \draw[->,very thick,red!70!black] ([xshift=-9mm]mpay.west) --
        node[above,font=\tiny,align=center]{\texttt{e\_entry}\\0x403000} (mpay.west);
      \draw[->,thick,dashed,black!60] (mpay.east)
        .. controls +(0.9,0) and +(0.9,0) ..
        node[right,font=\tiny,align=center,pos=0.5]{jumps to\\0x401050} (mrx.east);
    \end{tikzpicture}
    \caption{At run time the payload maps as a new executable segment, runs
      first, and jumps to the original entry.}
    \label{fig:elf-mem-after}
  \end{subfigure}
  \caption{How the injector rewrites the running example of
  Figure~\ref{fig:elf-memory}.  It \emph{redirects} \texttt{e\_entry} and
  \texttt{e\_shoff}, \emph{repurposes} the spare \texttt{PT\_NOTE} header into an
  executable \texttt{PT\_LOAD}, and \emph{appends} the payload (named by a
  \texttt{.payload} section) together with a new section-header table.
  The original program code and data are preserved.
  Execution begins in the appended segment and then returns to the host's original entry point.}
  \label{fig:elf}
\end{figure*}
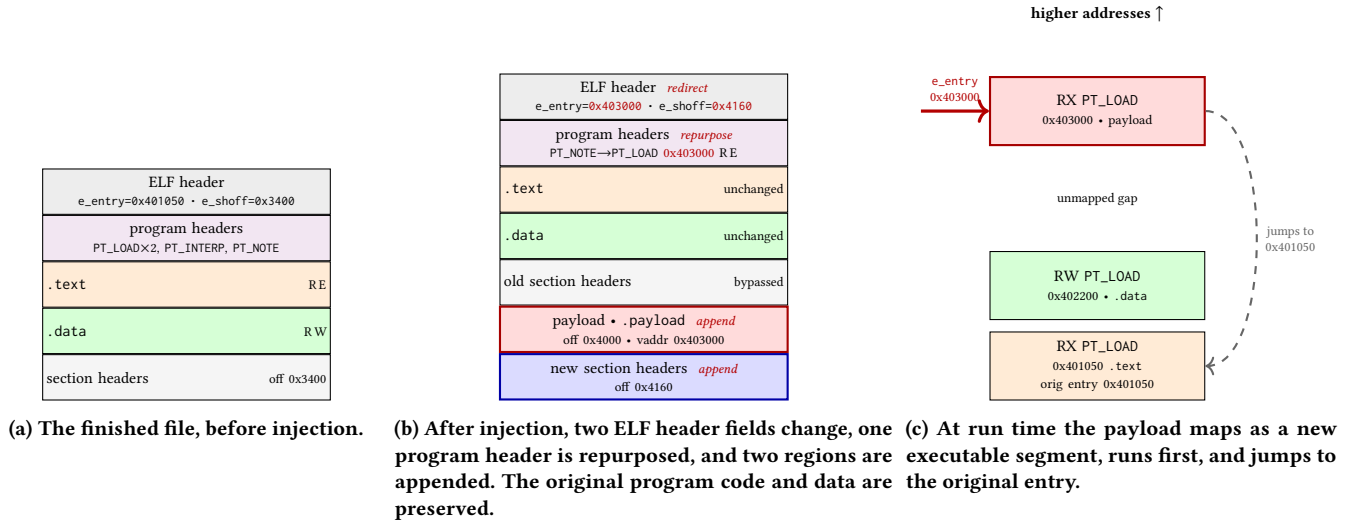

The produced ELF file must also stay coherent from the section-table view; otherwise, the nixpkgs post-processing step would discard the appended code.
The injector therefore describes the payload as a new \texttt{.payload} section.
It cannot enlarge the original section table in place, because the memory layout leaves no room for a larger table.
It instead builds a new section table at the end of the file.
This new table extends the original one with the \texttt{.payload} entry.
The injector then \emph{redirects} \texttt{e\_shoff} to the new table, so the new table overrides the original.
The old section table stays in the file but is now unreachable.
Figure~\ref{fig:elf} applies all three transformations to the running example of \Cref{fig:elf-memory}, showing the file before and after injection and the resulting memory layout.

\subsection{Propagation}
\label{sec:propagation}

The second stage of Figure~\ref{fig:overview} shows how the payload spreads through the bootstrap along the successor edge.
Each infected \texttt{strip} implants the payload into a newly built \texttt{strip}, which is then used by a later bootstrap stage.

Every package build compiles the package from source and then runs \texttt{strip} over the installed binaries during the fixup phase.
The upstream source and Nix recipe remain unmodified, but the build environment contains an infected \texttt{strip}.
Let $\texttt{strip}_i$ denote the copy of \texttt{strip} used at bootstrap stage~$i$, starting with the seed's trojaned $\texttt{strip}_0$.
When \texttt{strip} is rebuilt, the build phase compiles a fresh $\texttt{strip}_{1}$ from unmodified source, and the fixup phase runs $\texttt{strip}_0$ over the newly built binaries, including $\texttt{strip}_{1}$ itself.
Because $\texttt{strip}_0$ already carries the payload, it runs before \texttt{strip}'s own code, detects that it executes as \texttt{strip}, and uses the injector of \Cref{sec:seed-tampering} to implant the payload into $\texttt{strip}_{1}$ before stripping it.
So $\texttt{strip}_{1}$ is backdoored despite being built from unmodified source, and it implants the payload into $\texttt{strip}_{2}$ in turn.
The chain carries the payload from $\texttt{strip}_0$ to the \texttt{strip} of the final standard environment.

The compromised \texttt{strip} can also implant the payload into any eligible binary it fixes up.
Because each package runs a fixup phase by default, and that phase invokes the compromised \texttt{strip}, every binary the distribution produces carries the payload.

\subsection{Compromise}
\label{sec:compromise}

The third stage of Figure~\ref{fig:overview} shows the effect on the end user.
The payload runs whenever a backdoored binary starts, and on every host it executes its malicious code.

The payload must attach to something every executable has, so that it infects as many executables as possible, and it must run without any help from the host.
It therefore lives at the program's entry point, which the kernel reaches before any language runtime starts.
This constraint forces the payload to be self-contained, so we write it as a \emph{freestanding} program~\cite{cppreferenceFreestanding}.
Such a program links against no libraries, avoids libc and the host's dynamic symbols, and calls the kernel directly.
We need this because we implant the identical bytes into executables with unrelated dependencies, so the payload can rely on nothing they provide.

Because it runs even before the ordinary \texttt{\_start} routine, the payload is a careful guest.
It saves the registers the kernel handed to the process, aligns the stack as the x86-64 calling convention requires, and later restores the registers that \texttt{\_start} expects untouched.
It \emph{jumps} to the host's original entry point rather than forking or replacing the process, so the host keeps its PID, file descriptors, signal context, initial stack, and startup sequence.

\begin{figure}[t]
  \centering
  \begin{tikzpicture}[>=Latex,
      box/.style={draw,rounded corners,align=center,font=\small,minimum height=6.5mm,text width=44mm},
      entry/.style={box,fill=gray!13},
      proc/.style={box,fill=gray!6},
      dec/.style={diamond,draw,aspect=2.3,align=center,font=\small,inner sep=0pt,
                  fill=gray!16,text width=19mm,minimum height=9mm},
      side/.style={draw,rounded corners,align=center,font=\footnotesize,minimum height=7mm,text width=22mm},
      marker/.style={side,fill=gray!6},
      danger/.style={side,fill=red!16,draw=red!60!black,thick},
      lbl/.style={font=\scriptsize,fill=white,inner sep=1pt},
      flow/.style={->,thick}]
    \node[entry] (kernel) {\stepnum{1}~kernel enters implanted \texttt{e\_entry}};
    \node[proc,below=5mm of kernel] (save) {\stepnum{2}~save registers, align stack};
    \node[dec,below=6mm of save] (sandbox) {\stepnum{3}~inside Nix build?};
    \node[dec,below=16mm of sandbox] (host) {\stepnum{4}~basename is \texttt{strip}?};
    \node[entry,below=16mm of host] (return) {\stepnum{5}~restore state, jump to original \texttt{e\_entry}};
    \node[marker,right=12mm of sandbox] (marker) {emit marker};
    \node[danger,right=12mm of host] (inject) {infect \\eligible files};
    \draw[flow] (kernel) -- (save);
    \draw[flow] (save) -- (sandbox);
    \draw[flow] (sandbox) -- node[lbl,left]{yes} (host);
    \draw[flow] (sandbox) -- node[lbl,above]{no} (marker);
    \draw[flow] (marker) to[out=-90,in=20] (host.north east);
    \draw[flow] (host) -- node[lbl,left]{no} (return);
    \draw[flow] (host) -- node[lbl,above]{yes} (inject);
    \draw[flow,red!60!black] (inject) to[out=-90,in=20] (return.north east);
  \end{tikzpicture}
  \caption{Runtime control flow of an implanted executable. Every host runs the prologue, but only a host named \texttt{strip} infects the files passed on its command line. The circled numbers match the walk-through in the text.}
  \label{fig:runtime}
\end{figure}
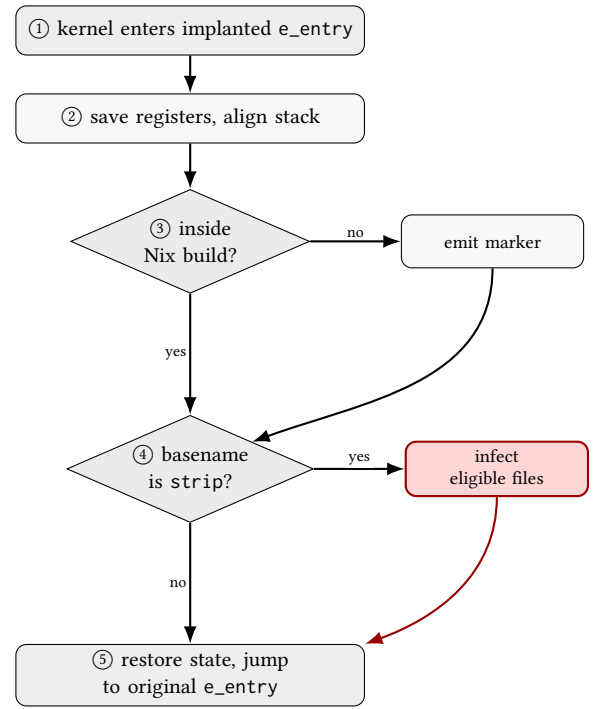

Figure~\ref{fig:runtime} traces the runtime path.
\stepnum{1}~The kernel enters the process at the implanted entry point, not the host's own.
\stepnum{2}~The payload saves the kernel's registers and aligns the stack.
\stepnum{3}~It checks for a Nix build, emitting the infection marker outside the sandbox and staying silent inside.
\stepnum{4}~It checks its own basename and, if it is \texttt{strip}, injects the payload into eligible files passed on the command line.
\stepnum{5}~It restores the saved state and jumps to the host's original entry point.

The payload detects a Nix build by reading the \texttt{NIX\_BUILD\_TOP} environment variable, and it stays silent inside the sandbox so that its behavior never fails a functional test during a build.
Outside the sandbox it executes its malicious code with the same permissions as the caller of the subverted executable.
In our prototype that behavior is a single fixed infection marker, the string \texttt{``You've been pwned!\textbackslash n''}, written to standard output before the host runs.
It stands in for any action the payload could take instead, such as stealing the user's credentials or tokens in the background.

\section{Experimental Evaluation}
\label{sec:evaluation}

We evaluate the attack along three axes.
First, does the compromised utility prevent any package from building?
Second, which downstream outputs does it reach?
Finally, does the payload injection impede the normal run-time behavior of subverted packages?

\subsection{Protocol}

We use nixpkgs at revision \texttt{fef9403a3e4d}, whose bootstrap uses GNU
binutils~2.44, and evaluate on \texttt{x86\_64-linux}.

We construct a compromised \texttt{strip} starting from the original one provided by nixpkgs.
We then build the graphical ISO installer, which covers a broad range of distribution software.
We count the number of packages from that package set that build correctly under our compromised seed.
We detect infected packages by running them and checking for console output containing the infection marker \texttt{``You've been pwned!\textbackslash n''}. Finally, we run the functional test suite of NixOS for the graphical image, to detect any behavioral degradation of the built packages.

\subsection{Results}

\subsubsection*{Buildability}

The trojaned seed does not affect the buildability of the package set, and the complete graphical ISO image builds successfully.
The runtime closure of the graphical installer, meaning the installer output and all store paths it references recursively, contains 1{,}199 package outputs and 3{,}799 user-invokable ELF executables.
Its total on-disk size is 6.16~GB and the median package is 0.39~MB.

\subsubsection*{Reach}

Out of 3{,}799 user-invokable binaries, 3{,}791 are CLI invokable.
Of these, \textbf{3{,}790} print the infection marker when invoked normally.
The single binary that does not print the marker is \texttt{firefox 147.0.3}. This is fully explained by the fact that the recipe passes \texttt{--disable-strip} \texttt{--disable-install-strip} to its build configuration, opting out of the strip phase entirely.

The graphical ISO image covers packages from multiple programming languages, confirming the language independence noted in \Cref{sec:objective}, since the attack acts on binaries directly.
The ten rows in Table~\ref{tab:ecosystems} present a non-exhaustive view of the diversity of software present in the graphical ISO package set. 

\begin{table}[t]
  \centering
  \caption{Selection of executables in the trojaned graphical installer closure, by source-language ecosystem. Every listed binary carries the infection marker.}
  \label{tab:ecosystems}
  \begin{tabular}{llrc}
    \toprule
    Ecosystem & Example binary & Size & Infected? \\
    \midrule
    C / C++  & \texttt{bash}                              & 1{,}327\,KB & \Checkmark \\
    C / C++  & \texttt{git}                               & 56{,}313\,KB & \Checkmark \\
    C / C++  & \texttt{sudo}                              &    248\,KB & \Checkmark \\
    C / C++  & \texttt{curl}                              &    222\,KB & \Checkmark \\
    C / C++  & \texttt{python3.13}                       & 10{,}928\,KB & \Checkmark \\
    Python   & \texttt{pydoc3.13}  &    110\,KB & \Checkmark \\
    Python   & \texttt{idle3.13}   &     61\,KB & \Checkmark \\
    Rust     & \texttt{rsvg-convert}                     & 8{,}065\,KB & \Checkmark \\
    Go       & \texttt{captree}                           & 1{,}964\,KB & \Checkmark \\
    Lua      & \texttt{lua}                               &    328\,KB & \Checkmark \\
    \bottomrule
  \end{tabular}
\end{table}

\subsubsection*{Runtime behavior}

We run the functional tests of the NixOS distributions associated with the graphical installer. The tests exercise several end-to-end user workflows by running a full NixOS virtual machine in a graphical environment and emulating real workloads like interacting with graphical applications, or installing a NixOS machine from scratch.
Again, every binary in the session VM, including \texttt{gnome-shell}, \texttt{mutter}, \texttt{nautilus}, and the GNOME control center, is built by the trojaned \texttt{strip}.
No test fails, and the infection payload does not break any program covered by the test suite.

\section{Discussion}
\label{sec:discussion}

\subsection{Limitations}

The current injector makes a few assumptions about the shape of the ELF file it works with.
It assumes a little-endian ELF file with a spare \texttt{PT\_NOTE} header and either an \texttt{ET\_EXEC} binary or an interpreted \texttt{ET\_DYN} one.
Shared libraries, other architectures, static \texttt{PIE} executables without an interpreter, and binaries with no suitable note segment fall outside its current scope.
These restrictions simplify the research prototype and are not fundamental limitations of the approach.

The implementation presented in this article may not generalize directly to Linux distributions other than NixOS because bootstrap procedures differ and may not use \texttt{strip} in the default build process.
However, the self-propagating pattern is more general than this implementation and applies to any distribution whose build graph contains the successor edge from \Cref{sec:pattern}.

Our prototype is also deliberately easy to detect because the visible marker, the \texttt{.payload} section name, and the basename check provide experimental instrumentation rather than conditions required by the propagation pattern.
We also make the attack propagate to \emph{every} package in the distribution to test the extent of its reach.
In practice, an attacker could keep the attack mostly dormant and propagate it only to future builds of \texttt{strip} until it reaches its ultimate target, possibly years later, which would make detection extremely difficult.

\subsection{Other Candidate Carriers}
\label{sec:generality-utils}
To identify other possible carriers, we examine the bootstrap utilities against the conditions in \Cref{sec:pattern}.
A utility must be trusted before it is built from source, modify a newly built copy of itself, and pass that role on to the modified copy.

We inspect the nine default build phases and their shared hooks in the evaluated nixpkgs revision.\footnote{\url{https://github.com/NixOS/nixpkgs/blob/fef9403a3e4d31b0a23f0bacebbec52c248fbb51/pkgs/stdenv/linux/bootstrap-files/x86_64-unknown-linux-gnu.nix}}
We find close to forty bootstrap utilities invoked directly by name and check these invocations by tracing a minimal package build.
This search excludes commands invoked indirectly through package Makefiles, which we discuss separately below.

Besides \texttt{strip}, \texttt{patchelf} meets all three conditions.
It runs during fixup to adjust runtime library paths in newly built executables.
When nixpkgs rebuilds \texttt{patchelf}, an earlier copy modifies the new one.
A compromised copy could therefore infect its replacement, which would then process later builds.

\texttt{install} and \texttt{cp} are also plausible carriers.
They copy newly built executables into their installation directories, giving them an opportunity to inject a payload.
In particular, rebuilding coreutils uses an earlier \texttt{install} to copy the newly built \texttt{install} binary.
The infected replacement could then propagate the payload during later installations.
Unlike \texttt{strip} and \texttt{patchelf}, these utilities depend on the package's installation commands to reach their targets.
We therefore consider \texttt{install} and \texttt{cp} plausible carriers, rather than established alternatives to \texttt{strip}.

The other utilities in \Cref{tab:seed-utils} do not directly rewrite finished executables during their usual build tasks.
For example, \texttt{make} runs build commands, \texttt{sed} and \texttt{patch} edit source text, and \texttt{tar} and \texttt{gzip} process archives.
They therefore lack the mechanism that lets \texttt{strip} infect both downstream programs and later copies of itself.

\begin{table}[t]
  \caption{Selected bootstrap utilities and the conditions of \Cref{sec:pattern}. C1 requires trust before building the utility from source. C2 requires an earlier copy to modify a newly built copy of itself. C3 requires the modified copy to perform the same operation in later builds. Parenthesized checkmarks indicate that the condition depends on the package's installation commands.}
  \label{tab:seed-utils}
  \centering
  \small
  \begin{tabular}{@{}llccc@{}}
    \toprule
    Utility & Phase & C1 & C2 & C3 \\
    \midrule
    \texttt{strip}    & fixup           & \Checkmark & \Checkmark & \Checkmark \\
    \texttt{patchelf} & fixup           & \Checkmark & \Checkmark & \Checkmark \\
    \midrule
    \texttt{install}, \texttt{cp} & install & \Checkmark & (\Checkmark) & (\Checkmark) \\
    \midrule
    \texttt{make}                               & build, install   & \Checkmark & --- & --- \\
    \texttt{sed}, \texttt{grep}, \texttt{patch} & patch, configure & \Checkmark & --- & --- \\
    \texttt{tar}, \texttt{gzip}, \texttt{xz}    & unpack, patch    & \Checkmark & --- & --- \\
    \texttt{ln}, \texttt{rm}, \texttt{mkdir}    & all              & \Checkmark & --- & --- \\
    \bottomrule
  \end{tabular}
\end{table}

\subsection{Alternative Architecture}
\label{sec:alternative}

We built the attack as an entry-point \emph{parasite}, a small payload injected into the genuine \texttt{strip}, but the same propagation also works with a \emph{wrapper} architecture.
In this version, the seed \texttt{strip} is replaced by a program that carries the genuine \texttt{strip} as embedded bytes, runs that copy at invocation time, and applies the payload to the files on its command line, rewriting each newly built copy into a fresh wrapper.
Because the wrapper runs as an ordinary program rather than a constrained pre-\texttt{main} hook, it gives the attacker more control.
It can act before, during, and after \texttt{strip} runs, rewrite the environment, carry any seed utility regardless of its ELF layout, and even replace \texttt{strip} with an unrelated program.
Our parasite deliberately performs one synchronous, self-contained action before returning control from a runtime in which libc and the host's normal initialization are unavailable.
More elaborate behavior remains possible, but it would require implementing additional runtime functionality directly and would increase compatibility risk.

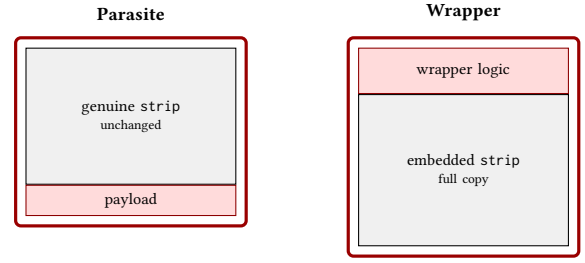
\begin{figure}[t]
  \centering
  \begin{tikzpicture}[
    >=Latex,
    reg/.style={draw,align=center,font=\scriptsize,text width=26mm,inner sep=2.5pt},
    genuine/.style={reg,fill=gray!12},
    mal/.style={reg,fill=red!14,draw=red!55!black},
    frame/.style={draw=red!60!black,very thick,rounded corners=2pt,inner sep=1.3mm},
    ttl/.style={font=\footnotesize\bfseries},
    dim/.style={font=\scriptsize,text=black!60}]
    \node[genuine,minimum height=18mm,anchor=north] (pg) at (0,0) {genuine \texttt{strip}\\{\tiny unchanged}};
    \node[mal,minimum height=4mm,below=0pt of pg] (pp) {payload};
    \node[frame,fit=(pg)(pp)] (pf) {};
    \node[ttl,above=1mm of pf.north] {Parasite};
    \node[mal,minimum height=6mm,anchor=north] (ww) at (4.4,0) {wrapper logic};
    \node[genuine,minimum height=20mm,below=0pt of ww] (we) {embedded \texttt{strip}\\{\tiny full copy}};
    \node[frame,fit=(ww)(we)] (wf) {};
    \node[ttl,above=1mm of wf.north] {Wrapper};
  \end{tikzpicture}
  \caption{Comparison of the two possible attack architectures. The \emph{parasite} leaves the genuine
  \texttt{strip} in place and appends a tiny entry-point payload.  The \emph{wrapper} instead replaces
  \texttt{strip} with a program that carries a full embedded copy of it. Gray marks
  genuine \texttt{strip} code and red marks the attacker's own additions.}
  \label{fig:arch-compare}
\end{figure}

We nonetheless use the parasite, because it is stealthier and less likely to break a build.
First, the parasite leaves the genuine \texttt{strip} untouched, so an infected binary still looks like \texttt{strip} to anyone who inspects it, whereas the wrapper replaces \texttt{strip} with a different program that embeds a whole copy of it and no longer resembles the tool it stands in for (\Cref{fig:arch-compare}).
Second, the parasite hands control to the genuine \texttt{strip} and inherits its behavior exactly, whereas the wrapper must faithfully reproduce how \texttt{strip} is invoked, and any discrepancy risks corrupting an output or failing a package build.

\section{Related Work}
\label{sec:related}

\subsection{The Trusting-Trust Attack and its Generalizations}

In his Turing Award lecture, Thompson~\cite{thompson1984trusting} describes a compiler he modified to insert a backdoor into selected programs and reproduce the attack when compiling itself.
The backdoor persists even after the malicious code is removed from the compiler's source.
Our attack follows the same principle, but uses \texttt{strip} to modify compiled binaries.
Whereas the compiler targets programs in the language it compiles, \texttt{strip} can infect eligible ELF binaries produced by different languages and toolchains.
\Cref{tab:comparison} compares the two implementations.

\begin{table}[t]
  \caption{How Thompson's compiler and our build utility fill the same four roles.}
  \label{tab:comparison}
  \centering
  \small
  \begin{tabular}{lll}
    \toprule
    Role & Thompson's compiler~\cite{thompson1984trusting} & This work \\
    \midrule
    Transformation & compiles source & strips ELF \\
    Recognition & source pattern & host basename \\
    Implantation & code generation & binary rewriting \\
    Successor edge & compiles compiler & strips \texttt{strip} \\
    \bottomrule
  \end{tabular}
\end{table}

Other work extends Thompson's idea beyond the original compiler.
Karger and Schell~\cite{karger1974multics} describe compiler-mediated code insertion in their Multics vulnerability analysis.
Spinellis~\cite{spinellis_reflections_2003a} shows that a platform's security policy is untrustworthy once an untrusted agent runs on it, analogous to how an untrusted compiler implies that its source is also untrusted.
Bratus et al.~\cite{bratus_planted_2014} argue that the essential ingredient is not a planted compiler bug but the way programs handle input, since any program that reads untrusted input can carry the same attack.
Maynor~\cite{maynor_trust_2004} names the idea but does not instantiate the attack, and observes only that the weak link generalizes beyond the compiler to other trusted build tools.
Our work supplies the missing instantiation.
We show that a post-build utility, rather than a compiler, sustains a complete trusting-trust chain through a real distribution bootstrap. We also give a concrete end-to-end implementation of such an attack.

\subsection{Verifying the Source-to-Binary Correspondence}

A parallel line of work tries to establish trust in a binary by tying it back to its source.
It targets the gap that our attack exploits, the divergence between the trusted source and the shipped binary.
Wheeler~\cite{wheeler2005ddc,wheeler2009ddc} introduces diverse double-compiling and supplies executable demonstrations, including a deliberately corrupted compiler and experiments with GCC.
Diverse double-compiling rebuilds a compiler with an independent second compiler.
It then checks that the two results agree.
A disagreement exposes a Thompson-style implant that only one of the two toolchains carries.
Reproducible builds pursue the same correspondence at distribution scale~\cite{lamb2022reproducible,11025777}.
They make every build produce bit-for-bit identical output.
An independent rebuilder then confirms that a binary matches its source.

Both techniques secure the path from source to binary, but neither inspects the utilities that run after the compiler finishes.
Our attack lives in exactly that gap.
It hides inside \texttt{strip}, which the original build and every rebuild trust alike.
Diverse double-compiling diversifies the compiler and not a build utility like \texttt{strip}, so the tampered utility sits on both sides of the comparison and cancels out.
Reproducible builds confirm that a binary is deterministic and not that its build environment is honest, so a rebuild that reuses the same seed reproduces the same implant bit-for-bit.
The recipe, the dependency records, and the source all stay intact, so neither check fires.

\subsection{Bootstrappable Builds}

The Bootstrappable Builds project~\cite{bootstrappable} addresses the same problem at its root by shrinking the set of opaque binaries a distribution must trust.
The full-source bootstrap in GNU Guix~\cite{guixFullSourceBootstrap} reduces the binary seed to a few hundred bytes and rebuilds the whole toolchain from auditable source above it~\cite{gnuMes}.
Camlboot~\cite{courant2022camlboot} performs an analogous full-source bootstrap for the OCaml toolchain by stacking simple implementations and checking the result with diverse compilation.

\subsection{Build-Time and Distribution Supply-Chain Attacks}

Documented incidents show attacks against build and distribution pipelines rather than source repositories alone~\cite{crowdstrikeSunspot,freundXz,hal-05326226}.
The SolarWinds compromise inserts its backdoor with SUNSPOT, an implant that watches for the compiler and rewrites a source file during the build, so the shipped binary diverges from the repository~\cite{crowdstrikeSunspot}.
The xz and liblzma backdoor hides its payload in build fixtures and activates it through the release build machinery, again placing malicious logic in the build rather than in reviewable source~\cite{freundXz, hal-05326226}.
Ohm and colleagues catalogue hundreds of malicious open-source packages and organize their injection and trigger techniques into attack trees~\cite{ohm2020backstabber}.
Defensive frameworks such as in-toto attest that each build step ran as intended and that artifacts pass unmodified between steps~\cite{torresarias2019intoto}.
The Shai-Hulud worm~\cite{shaihulud} shows that attackers now pursue propagation directly.
It steals a maintainer's credentials and republishes trojanized versions of the other npm packages that the victim controls, so one compromise spreads across the registry.
This propagation is horizontal and source-level.
Each infected package still ships malicious, reviewable source, and rebuilding that package from unmodified upstream source removes the infection.
Most other incidents concern a single compromised build or package, and even Shai-Hulud propagates through reviewable source.
Our attack is instead self-propagating through the build itself.
One tampered seed reproduces the implant in later \texttt{strip} binaries built from unmodified source.
The final \texttt{stdenv} remains infected without retaining a runtime reference to the seed, which neither source review nor a per-artifact attestation of an honestly executed build step detects.

\section{Conclusion}

We have shown that the trusting-trust attack is not limited to compilers.
GNU \texttt{strip}, an ordinary build utility that never inspects or generates source, sustains the whole attack using only manipulations of finished ELF files.
The implant propagates through one NixOS bootstrap execution, and the \texttt{strip} in the final standard environment remains infected without retaining a runtime reference to the malicious seed.
On a real nixpkgs revision it reaches almost every binary of a complete graphical installer across a diversity of language ecosystems, and breaks no build or functional test.

\balance
\bibliographystyle{ACM-Reference-Format}
\bibliography{main}

\end{document}